\documentclass[twocolumn,prb]{revtex4-1}
\usepackage{amsmath,amsbsy,amssymb,graphicx}
\usepackage{times}

\begin{document}

\title{{\Large Complex Structure of Triangular Graphene: }\\
{\Large Electronic, Magnetic and Electromechanical Properties}}
\author{Motohiko Ezawa}
\affiliation{Department of Applied Physics, University of Tokyo, Hongo 7-3-1, 113-8656,
Japan }

\begin{abstract}
We have investigated electronic and magnetic properties of graphene
nanodisks (nanosize triangular graphene) as well as electromechanical
properties of graphene nanojunctions. Nanodisks are nanomagnets made of
graphene, which are robust against perturbation such as impurities and
lattice defects, where the ferromagnetic order is assured by Lieb's theorem.
We can generate a spin current by spin filter, and manipulate it by a spin
valve, a spin switch and other spintronic devices made of graphene
nanodisks. We have analyzed nanodisk arrays, which have multi-degenerate
perfect flat bands and are ferromagnet. By connecting two triangular
graphene corners, we propose a nanomechanical switch and a rotator, which can detect
a tiny angle rotation by measuring currents between the two corners. By
making use of the strain induced Peierls transition of zigzag nanoribbons, we also 
propose a nanomechanical stretch sensor, in which the conductance can be switch
off by a nanometer scale stretching. \newline
\end{abstract}

\maketitle

\section{Introduction}

Graphene, which is a one-layer thick honeycomb structure of carbon, is an
amazing material\cite{GraphEx}. Electrons exhibit high mobility and travel
micron distances without scattering at room temperature. It is a very thin
and strong material, showing very high thermal conductivity. Graphene is now
a main topic of nanoscience.

Much attention has been focused on graphene nanoribbons, which is a
one-dimensional ribbon-like derivatives of graphene. They have various band
structure depending on the edge and width. In particular, zigzag gaphene
nanoribbons show edge ferromagnetism due to almost flat low-energy band at
the Fermi level. There are a profusion of papers on them, among which we
cite some of early works\cite{Fujita,EzawaRibbon,Brey}. Another basic
element of graphene derivatives is a graphene nanodisk\cite{EzawaDisk}. It
is a nanometer-scale disk-like material which has a closed edge. It may be
considered as a giant molecule made of aromatic compound. It is possible to
manufacture them by etching a graphene sheet by Ni nanoparticles\cite%
{Herrero}. Among them, trigonal zigzag nanodisks have a novel electric
property that there exist half-filled zero-energy states in the
non-interacting regime, as was revealed first by the tight-binding model\cite%
{EzawaDisk} and then by first-principle calculations\cite%
{Fernandez,Hod,WangN}. Various remarkable properties of nanodisks have been
investigated extensively in a series of works\cite%
{EzawaDisk,EzawaCouloKondo,EzawaSpin,EzawaDirac}. Nanodisk is also referred
to as nanoisland\cite{Fernandez}, nanoflake\cite{WangN,WangL,Akola},
nanofragment\cite{YazyevROP} or graphene quantum dot\cite{GucluL,GucluB}.

Nanoribbons and nanodisks correspond to quantum wires and quantum dots,
respectively. They are candidates of future carbon-based nanoelectronics and
spintronics alternative to silicon devices. A nanoribbon-nanodisk complex
can in principle be fabricated, embodying various functions, only by etching
a graphene sheet. Furthermore, graphene is common material and ecological.
In this paper, exploring electronic, magnetic and electromechanical
properties of trigonal zigzag graphene nanodisks [see Fig.1], we propose
some application of nanodisk-nanoribbon complex to nanoelectronics,
spintronics and electromechanics devices

\begin{figure}[t]
\centerline{\includegraphics[width=0.40\textwidth]{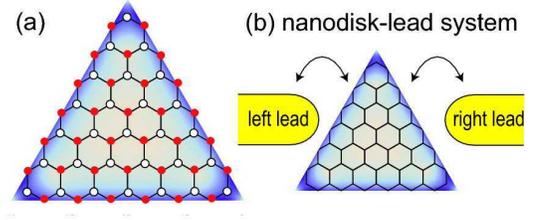}} \label%
{FigNLeadY}
\caption{
(a) Geometric configuration of a trigonal zigzag
nanodisk. The nanodisk size is defined by $N=N_{\text{ben}}-1$ with $N_{%
\text{ben}}$\ the number of benzenes on one side of the trigon. Here, $N=5$.
The A (B) sites on the lattice are indicated by red dots (while circles).
The electron density is found to be localized along the edges. \ (b)\ \ The
nanodisk-lead system. A nanodisk is connected to the right and left leads by
tunneling coupling. It may act as a spin filter.
}
\end{figure}

\section{Energy spectrum}

We calculate the energy spectrum of the nanodisk based on the
nearest-neighbor tight-binding model, which has proved to describe
accurately the electronic structure of graphene, carbon nanotubes, graphene
nanoribbons and other sp$^{2}$ carbon materials. The Hamiltonian is defined
by%
\begin{equation}
H_{0}=\sum_{i}\varepsilon _{i}c_{i}^{\dagger }c_{i}+\sum_{\left\langle
i,j\right\rangle }t_{ij}c_{i}^{\dagger }c_{j},  \label{HamilTB}
\end{equation}%
where $\varepsilon _{i}$ is the site energy, $t_{ij}$ is the transfer
energy, and $c_{i}^{\dagger }$ is the creation operator of the $\pi $
electron at the site $i$. The summation is taken over all nearest
neighboring sites $\left\langle i,j\right\rangle $. Owing to their
homogeneous geometrical configuration, we may take constant values for these
energies, $\varepsilon _{i}=\varepsilon _{\text{F}}$ and $t_{ij}=t\approx
2.70$eV. There exists one electron per one carbon, and the band-filling
factor is 1/2. Then, the diagonal term yields just a constant, $\varepsilon
_{\text{F}}N_{\text{C}}$, and can be neglected in the Hamiltonian, where $N_{%
\text{C}}$ is the number of carbon atoms.

We define the size $N$ of a nanodisk by $N=N_{\text{ben}}-1$, where $N_{%
\text{ben}}$ is the number of benzenes on one side of the trigon as in
Fig.1(a). It can be shown\cite{EzawaDisk} that the determinant associated
with the Hamiltonian (\ref{HamilTB})\ has such a factor as 
\begin{equation}
\det \left[ \varepsilon I-H\left( N_{\text{C}}\right) \right] \propto
\varepsilon ^{N},
\end{equation}%
implying $N$-fold degeneracy of the zero-energy states. The gap energy is as
large as a few eV for nanodisks with small $N$, where it is a good
approximation to investigate the electron-electron interaction physics only
in the zero-energy sector, by projecting the system to the subspace made of
those zero-energy states. As we shall see in Section \ref{SecLargeNanodisk},
the approximation remains to be good even for those with large $N$.

\section{Trigonal Symmetry}

The symmetry group of a trigonal nanodisk is $C_{3v}$, which is generated by
the $2\pi /3$ rotation $\mathfrak{c}_{3}$ and the mirror reflection $\sigma
_{\mathfrak{v}}$. It has the representation \{$A_{1}$, $A_{2}$, $E$\}. The $%
A_{1}$ representation is invariant under the rotation $\mathfrak{c}_{3}$ and
the mirror reflection $\sigma _{\mathfrak{v}}$. The $A_{2}$ representation
is invariant under $\mathfrak{c}_{3}$ and antisymmetric under $\sigma _{%
\mathfrak{v}}$. The $E$ representation acquires $\pm 2\pi /3$ phase shift
under the $2\pi /3$ rotation. The $A_{1}$ and $A_{2}$ are 1-dimensional
representations (singlets) and the $E$ is a 2-dimensitional representation
(doublet). These properties are summarized in the following character table:%
\begin{equation}
\begin{array}{c|ccc}
\hline
C_{3v} & \mathfrak{e} & 2\mathfrak{c}_{3} & 3\sigma _{\mathfrak{v}} \\ \hline
A_{1} & 1 & 1 & 1 \\ 
A_{2} & 1 & 1 & -1 \\ 
E & 2 & -1 & 0 \\ \hline
\end{array}
\label{TrigoSymme}
\end{equation}

The zero-energy sector consists of $N$ orthonormal states. We are able to
index them by the wave number $k$ along the edge. It is a continuous
parameter for an infinitely long graphene edge. According to the
tight-binding-model result, the flat band emerges for%
\begin{equation}
-\pi \leq ak<-2\pi /3\quad \text{and}\quad 2\pi /3<ak\leq \pi .
\label{RegionK}
\end{equation}%
We focus on the wave function $\psi \left( x,y\right) $ at one of the A
sites on an edge, and investigate the phase shift when we step over to the
neighboring site [see Fig.1(a)]. There are $N$ links along one edge of the
size-$N$ nanodisk, for which we obtain the phase shift $Nak$, where $a$ is
the spacing between the neighboring A sites and $k$ is the wave number along
the edge. On the other hand, the phase shift is $\pi $ at the corner. The
total phase shift is $3Nak+3\pi $, when we encircle the nanodisk once. By
requiring the single-valueness of the wave function, it is found to be
quantized as%
\begin{equation}
ak_{n}=\pm \left[ (2n+1)/3N+2/3\right] \pi ,  \label{ValueK}
\end{equation}%
where it follows that $0\leq n\leq (N-1)/2$ from the allowed region (\ref%
{RegionK}) of the wave number.

We may group the states according to the trigonal symmetry (\ref{TrigoSymme}%
). With respect to the rotation there are three elements $\mathfrak{c}%
_{3}^{0}$, $\mathfrak{c}_{3}$, $\mathfrak{c}_{3}^{2}$, which correspond to $%
1 $, $e^{2\pi i/3}$, $e^{4\pi i/3}$. Accordingly, the phase shift of one
edge is $0$, $2\pi /3$, $4\pi /3$. The state is grouped according to the
representation of the trigonal symmetry group $C_{3v}$ as follows,%
\begin{equation}
\begin{array}{cc}
\left. 
\begin{array}{cc}
A_{1}\text{ (singlet)}: & |k_{n}^{0}\rangle +|-k_{n}^{0}\rangle , \\ 
A_{2}\text{ (singlet)}: & |k_{n}^{0}\rangle -|-k_{n}^{0}\rangle ,%
\end{array}%
\right\} & \text{for\quad }\displaystyle k_{n}^{0}=\frac{6n+3}{3Na}\pi , \\ 
\begin{array}{cc}
E\text{ (doublet)}: & |k_{n}^{\pm }\rangle ,\quad |-k_{n}^{\pm }\rangle ,%
\end{array}
& \text{for\quad }\displaystyle k_{n}^{\pm }=\frac{6n\pm 1}{3Na}\pi .%
\end{array}
\label{TrigoWave}
\end{equation}%
The zero-energy state is indexed by the quantized wave number as $%
|k_{n}^{\alpha }\rangle $ with (\ref{TrigoWave}).

To see the meaning of the wave number $k_{n}^{\alpha }$ more in detail\cite%
{EzawaDirac}, we have calculated the probability density flow,%
\begin{equation}
\mathcal{A}_{i}(x,y)=-i\psi ^{\ast }(x,y)\partial _{i}\psi (x,y)
\end{equation}%
for states $|k_{n}^{\alpha }\rangle $, which we show for the case of $N=7$
in Fig.2. We observe clearly a texture of vortices. These vortices manifest
themselves as magnetic vortices perpendicular to the nanodisk plane when the
electromagnetic fields are coupled.

\begin{figure}[t]
\centerline{\includegraphics[width=0.48\textwidth]{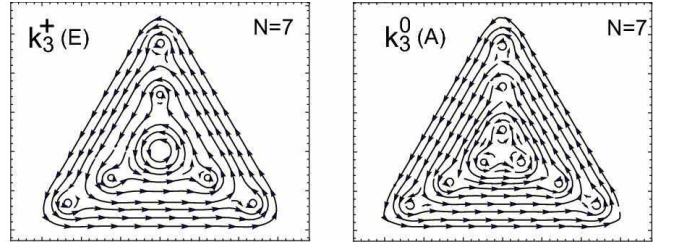}} \label%
{FigBerryN7}
\caption{
Probability density flow for
the zero-energy states in the nanodisk with $N=7$. The representation of the
trigonal symmetry group $C_{3v}$ is indicated in the parenthesis. A vortex
appears at the center of mass for the state belongs to the E (doublet)
representation. The winding number at the center is 2 in the state $%
|k_{n}^{+}\rangle $.
}\end{figure}

The total winding number $N_{\text{vortex}}$ is calculated by%
\begin{equation}
N_{\text{vortex}}=\frac{1}{2\pi }\oint dx_{i}\,\frac{\mathcal{A}_{i}(x,y)}{%
|\psi (x,y)|^{2}}=N+m-1,  \label{PhaseAB}
\end{equation}%
with $m=0,1,2,\cdots ,\left\lfloor (N-1)/2\right\rfloor $ in the size-$N$
nanodisk, where $\left\lfloor a\right\rfloor $ denotes the maximum integer
equal to or smaller than $a$. We find $N_{\text{vortex}}=3n$ for $k_{n}^{0}$%
, $N_{\text{vortex}}=3n+1$ for $k_{n+1}^{-}$ and $N_{\text{vortex}}=3n+2$
for $k_{n}^{+}$. The wave functions are classified in terms of modulo of the
total winding number: The wave function belongs to the E-representation and
has chiral edge mode for $N_{\text{vortex}}\equiv 1,2$ (mod 3), and belongs
to the A-representation and has non-chiral edge mode for $N_{\text{vortex}%
}\equiv 0$ (mod 3). The winding number of the vortex at the center of the
nanodisk is $0,1,2$ in the state $|k_{n}^{0}\rangle ,|k_{n}^{-}\rangle
,|k_{n}^{+}\rangle $, respectively.

\section{Quasiferromagnet}

The total spin of the ground state is determined by Lieb's theorem. The
total spin is given by the difference of the A site and the B site, $S=\frac{%
1}{2}\left\vert N_{A}-N_{B}\right\vert $, where $N_{A}$ and $N_{B}$ are
number of A site and B site. Here, $S=\frac{1}{2}N$ in the size-$N$
nanodisk, where $N_{A}=(N+1)(N+6)/2$ and $N_{B}=(N+2)(N+3)/2$. Hence we
expect a nanodisk to act as a ferromagnet. The ferromagnetic ground state is
robust against perturbations such as randomness and lattice defects since it
is assured by Lieb's theorem. This feature brings out a remarkable contrast
between nanodisks and nanoribbons. The ferromagnetic order is fragile due to
the lack of Lieb's theorem in the case of nanoribbons.

We investigate ferromagnetic properties by introducing Coulomb interactions
into the zero-energy sector\cite{EzawaDisk}. We have calculated specific
heats and susceptibilites at temperature $T$ in Fig.3. There appear
singularities in thermodynamical quantities as $N\rightarrow \infty $, which
represent a phase transition at $T_{c}$ between the ferromagnet and
paramagnet states, where $T_{c}=JN/2k_{\text{B}}$. For finite $N$, there are
steep changes around $T_{c}$, though they are not singularities. It is not a
phase transition. However, it would be reasonable to call it a quasi-phase
transition between the quasiferromagnet and paramagnet states. Such a
quasi-phase transition is manifest even in finite systems with $N=100$ $\sim 
$ $1000$.

\begin{figure}[t]
\centerline{\includegraphics[width=0.48\textwidth]{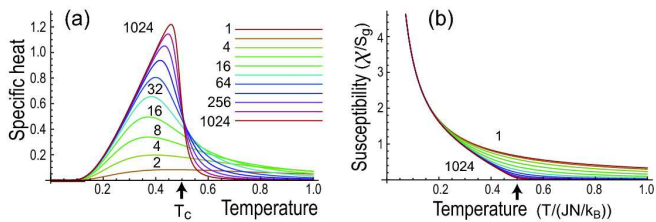}} \label{FigThermo}
\caption{
Thermodynamical properties of
the nanodisk-spin system. (a) The specific heat $C$ in unit of $k_{\text{B}%
}N $. (b) The susceptibility $\chi $ in unit of $S_{g}$. The size is $%
N=1,2,2^{2},\cdots 2^{10}$. The horizontal axis stands for the temperature $%
T $ in unit of $JN/k_{\text{B}}$. The arrow represents the phase transition
point $T_{c}$ in the limit $N\rightarrow \infty $.}
\end{figure}

The specific heat takes nonzero-value for $T>T_{c}$, as shown in Fig.3(a),
which is zero in the limit $N\rightarrow \infty $. The result indicates the
existence of some correlations in the paramagnet state. On the other hand,
the susceptibility $\chi $ always shows the Curie-Weiss low $\chi \propto
1/T $ near $T=0$, and exhibits also a behavior showing a quasi-phase
transition at $T=T_{c}$, as shown in Fig.3(b). In the finite system, the
expectation value of $S_{z,\text{tot}}$ is always zero because there is no
spontaneous symmetry breakdown in the finite system, and the behavior is
that of paramagnet.

\section{Magnetism of large nanodisks}

\label{SecLargeNanodisk}

For large $N$ nanodisks, the band gap decreases inversely proportional to
the size. One may wonder if our analysis based only on the zero-energy
sector is valid. Indeed, the size of experimentally available nanodisks is
as large as $N=100\sim 1000$. We wish to argue that our analysis based on
the zero-energy sector is essentially correct, even if the size $N$ of the
nanodisk is large and the band gap becomes very narrow.

Near the Fermi energy, the density of states (DOS) $D\left( \varepsilon
\right) $ consists of that of the bulk graphene and an additional peak at
the zero-energy states due to the edge states for $N\gg 1$, as illustrated
in Fig.4. Hence, together with spin degrees of freedom, it behaves as%
\begin{equation}
D\left( \varepsilon \right) =2cN_{\text{C}}\left\vert \varepsilon
\right\vert +2N\delta \left( \varepsilon \right) ,  \label{LargeDOS}
\end{equation}%
with a certain constant factor $c$. The linear term is due to the bulk
states, and the Dirac delta function term is due to the edge states. The
important point is that the edge-state peak is clearly distinguished from
the DOS due to the bulk part. It is enough to take into account only the
zero-energy sector to analyze physics near the Fermi energy, since the
contribution from the edge states is dominant.

\begin{figure}[t]
\centerline{\includegraphics[width=0.48\textwidth]{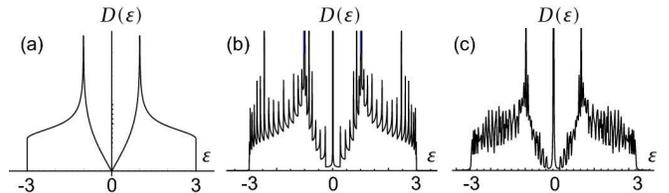}} \label%
{FigGrapDOS}
\caption{
The density of states of (a)
graphene, (b) graphene nanoribbon and (c) graphene nanodisk. The horizontal
axis is the energy $\varepsilon $ in unit of $t$.
}
\end{figure}

We calculate the magnetization of a nanodisk when its size is large. We
start with the Hubbard Hamiltonian,%
\begin{equation}
H=\sum_{k\sigma }\varepsilon \left( k\right) c_{k\sigma }^{\dagger
}c_{k\sigma }+U\sum_{kk^{\prime }q}c_{k+q\uparrow }^{\dagger }c_{k^{\prime
}-q\downarrow }^{\dagger }c_{k^{\prime }\downarrow }c_{k\uparrow }.
\end{equation}%
Let $\left\langle n_{\uparrow }\right\rangle ,\left\langle n_{\downarrow
}\right\rangle $ be the average numbers of the up and down spins. The
magnetization is given by $\left\langle m\right\rangle =\left\langle
n_{\uparrow }\right\rangle -\left\langle n_{\downarrow }\right\rangle $. It
is determined self-consistently by the relation%
\begin{equation}
\left\langle m\right\rangle =\int d\varepsilon \,D\left( \varepsilon \right) 
\left[ f\left( \varepsilon -\Delta \right) -f\left( \varepsilon +\Delta
\right) \right] ,  \label{StornerM}
\end{equation}%
in terms of the Fermi distribution function 
\begin{equation}
f\left( x\right) =\left( \exp [\left( x-\mu \right) /k_{\text{B}}T]+1\right)
^{-1}
\end{equation}%
and 
\begin{equation}
\Delta =\frac{U}{2N_{\text{C}}}\left\langle m\right\rangle +\frac{1}{2}h.
\end{equation}%
Substituting the formula (\ref{LargeDOS}) into the Storner equation (\ref%
{StornerM}), we obtain%
\begin{equation}
\left\langle m\right\rangle =N\tanh \frac{\beta \Delta }{2}+cN_{\text{c}}%
\left[ \Delta ^{2}+\frac{1}{\beta ^{2}}\left\{ \frac{\pi ^{2}}{3}+4\text{Li}%
_{2}\left( -e^{-\beta \Delta }\right) \right\} \right] ,
\end{equation}%
with the dilogarithm function Li$_{2}\left( x\right) $. It is difficult to
solve this equation for $\left\langle m\right\rangle $ self-consistently at
general temperature $T$. We examine two limits, $T\rightarrow 0$ and $%
T\rightarrow \infty $.

For the zero temperature ($T\rightarrow 0$) we obtain the magnetization as%
\begin{equation}
\left\langle m\right\rangle =N+c\frac{U^{2}}{4N_{\text{C}}}\left\langle
m\right\rangle ^{2}+\frac{cUh}{2}\left\langle m\right\rangle +O(h^{2}).
\label{MagneLargeN}
\end{equation}%
Because $\left\vert \left\langle m\right\rangle \right\vert \leq N$, it
follows that $\left\langle m\right\rangle =N+O(1)$. The contribution from
the bulk gives a negligible correction to the total magnetization. Hence the
magnetization is $\left\langle m\right\rangle =N$, and the ground state is
fully poralized whenever $U\neq 0$. Ferromagnetism occurs irrespective of
the strength of the Coulomb interaction. The magnetization is propotional
not to $N_{\text{C}}$ but $N$. In this sence the ground state of nanodisk is
not bulk ferromagnet but surface ferromagnet, which is consistent with the
previous result.

We next investigate the high temperatuer limit ($T\rightarrow \infty $).
Using the Taylor expansion of the dilogarithm function,%
\begin{equation}
\text{Li}_{2}\left( -e^{-\beta \Delta }\right) =-\frac{\pi ^{2}}{12}+\beta
\Delta \log 2-\frac{\beta ^{2}\Delta ^{2}}{4}+\frac{\beta ^{3}\Delta ^{3}}{24%
}+\cdots ,
\end{equation}%
we find%
\begin{equation}
\left\langle m\right\rangle =N\tanh \frac{\beta \Delta }{2}+cN_{\text{c}}%
\left[ \frac{4\Delta }{\beta }\log 2+\frac{\beta \Delta ^{3}}{6}+\cdots %
\right] .
\end{equation}%
The leading term is the second term, and hence the main contribution comes
from the bulk. The solution is only $\left\langle m\right\rangle =0$ for
which $\Delta =0$. There is no magnetization at high temperature.

A comment is in order. We have assumed that the magnetization axis is fixed
and only longitudinal fluctuations of the magnetic moments take place. In
general, spin-wave-like fluctuations are dominant at the edges of graphene%
\cite{YazyevL}, because they are gapless Goldstone modes. On the contrary,
there exist no gapless Goldstone modes in graphene nanodisks, because the
edge is finite and closed. Furthermore, when the length of the edge is very
small, spin-wave-like fluctuations have a large gap. Hence, our
approximation is valid for nanodisks.

\section{Application for spintronic devices}

\begin{figure}[t]
\centerline{\includegraphics[width=0.40\textwidth]{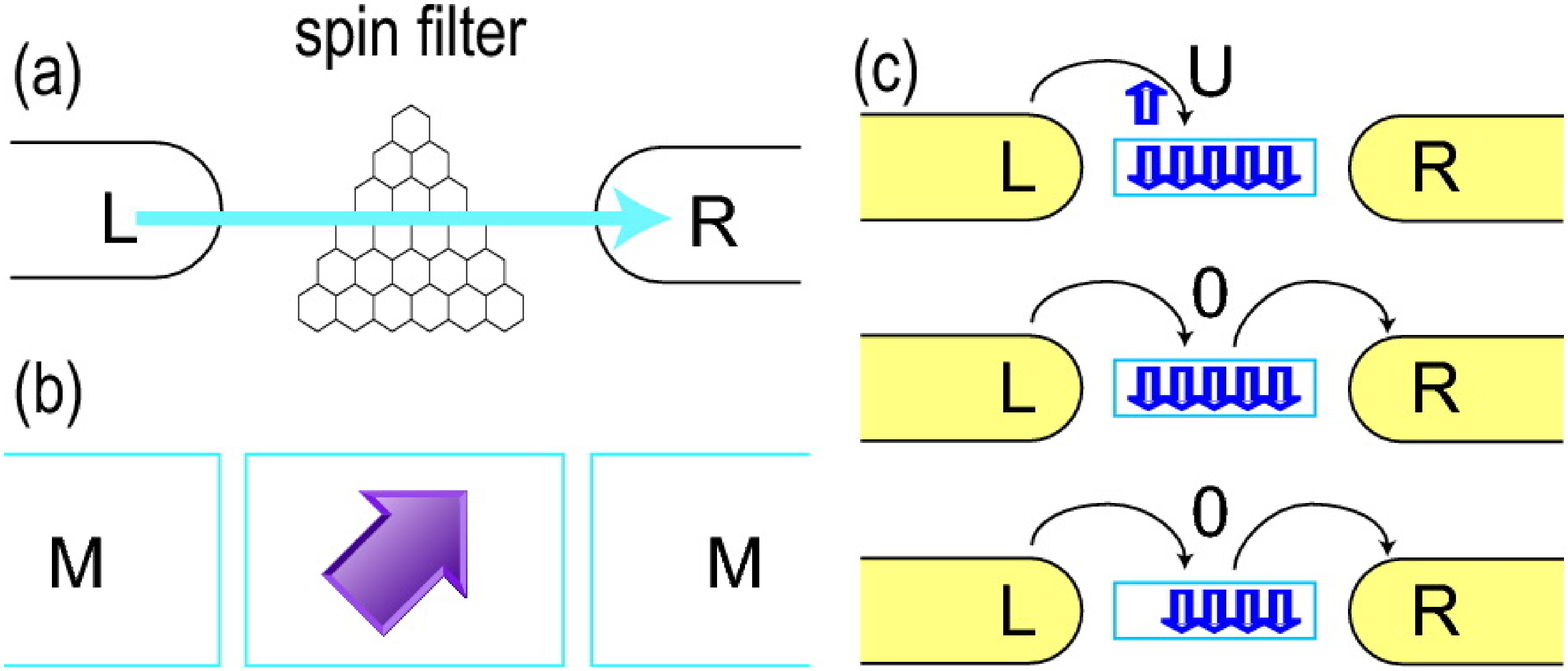}} \label{FigMFM}
\caption{
(a) An electron tunnels from
the left lead to the nanodisk and then to the right lead. The system is a
reminiscence of a metal-ferromagnet-metal junction. (b) Only electrons with
the same spin direction as the nanodisk spin can pass through the nanodisk
freely. As a result, when we apply a spin-unpolarized current to the
nanodisk, the outgoing current is spin polarized to the direction of the
nanodisk spin. Consequently, this system acts as a spin filter.}
\end{figure}

The nanodisk-spin system is a quasiferromagnet, which is an interpolating
system between a single spin and a ferromagnet. It is easy to control a
single spin by a tiny current but it does not hold the spin direction for a
long time. On the other hand, a ferromagnet is very stable, but it is hard
to control the spin direction by a tiny current. A nanodisk quasiferromagnet
has an intermediate nature: It can be controlled by a relatively tiny
current and yet holds the spin direction for quite a long time\cite%
{EzawaDisk}. Taking advantage of these properties we have already proposed
elsewhere\cite{EzawaSpin} some applications of graphene nanodisk-lead
systems to spintronic devices. They are spin filter, spin memory, spin
amplifier, spin valve, spin-field-effect transistor, spin diode and spin
switch, among which here we make a consice review of spin filter, spin valve
and spin switch. We newly propose nanodisk arrays and nanomechanical switch.

\textbf{Spin filter:} We consider a lead-nanodisk-lead system [see
Fig.1(b)], where an electron makes a tunnelling from the left lead to the
nanodisk and then to the right lead. This system is a reminiscence of a
metal-ferromagnet-metal junction [see Fig.5]. If electrons in the lead has
the same spin direction as the nanodisk spin, they can pass through the
nanodisk freely. However, those with the opposite direction feel a large
Coulomb barrier and are blocked (Pauli blockade [Fig.5(c)])\cite{EzawaSpin}.
As a result, when we apply a spin-unpolarized current to the nanodisk, the
outgoing current is spin polarized to the direction of the nanodisk spin.
Consequently, this system acts as a spin filter.

\textbf{Spin valve:} A nanodisk can be used as a spin valve, inducing the
giant magnetoresistance effect. We set up a system composed of two nanodisks
sequentially connected with leads [see Fig.6]. We apply external magnetic
field, and control the spin direction of the first nanodisk to be $%
\left\vert \theta \right\rangle =\cos \frac{\theta }{2}\left\vert \uparrow
\right\rangle +\sin \frac{\theta }{2}\left\vert \downarrow \right\rangle $,
and that of the second nanodisk to be $\left\vert 0\right\rangle =\left\vert
\uparrow \right\rangle $. We inject an unpolarized-spin current to the first
nanodisk. The spin of the lead between the two nanodisks is polarized into
the direction of $\left\vert \theta \right\rangle $. Subsequently the
current is filtered to the up-spin one by the second nanodisk. The outgoing
current from the second nanodisk is $I_{\uparrow }^{\text{out}}=I\cos \frac{%
\theta }{2}$. We can control the magnitude of the up-polarized current from $%
0$ to $I$ by rotating the external magnetic field. The system act as a spin
valve.

\begin{figure}[t]
\centerline{\includegraphics[width=0.40\textwidth]{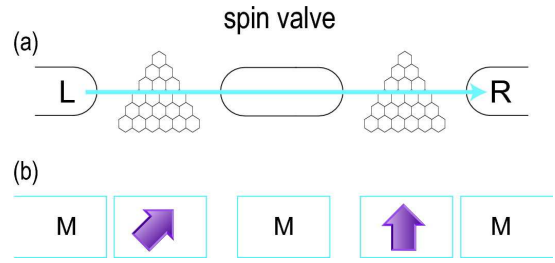}} \label%
{FigSpinValve}
\caption{
The spin valve is made of two
nanodisks with the same size, which are connected to leads. Applying an
external magnetic field, we control the spin direction of the first nanodisk
to be $\left\vert \theta \right\rangle $ and that of the second nanodisk to
be $\left\vert \uparrow \right\rangle $. The incoming current is
unpolarized, but the outgoing current is polarized and its magnitude can be
controlled continuously. This acts as a spin valve.
}
\end{figure}

\textbf{Spin switch:} We consider a chain of nanodisks and leads connected
sequentially [see Fig.7]. Without external magnetic field, nanodisk spins
are oriented randomly due to thermal fluctuations, and a current cannot go
through the chain. However, when and only when a uniform magnetic field is
applied to all nanodisks, the direction of all nanodisk spins become
identical and a current can go through. Thus the system acts as a spin
switch, showing a giant magnetoresistance effect. The advantage of this
system is that a detailed control of magnetic field is not necessary in each
nanodisk.

\begin{figure}[t]
\centerline{\includegraphics[width=0.46\textwidth]{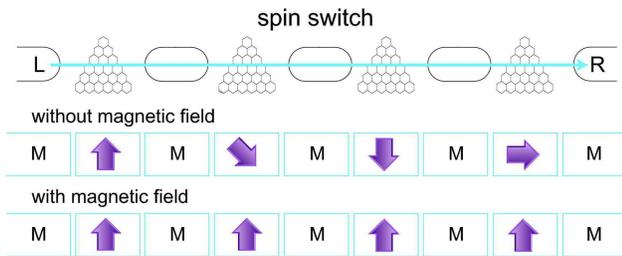}} \label{FigGMR}
\caption{
A chain of nonodisks and leads
acts as a spin switch. Without external magnetic field, nanodisk spins are
oriented randomly due to thermal fluctuations, and a current cannot go
through the chain. However, as soon as a uniform magnetic field is applied
to all nanodisks, the direction of all nanodisk spins become identical and a
current can go through.}
\end{figure}

\textbf{Nanodisk arrays:} We investigate nanodisk arrays, which are
materials where nanodisks are connected in one- or two-dimentions. These
structures have already been manufactured by etching a graphene sheet by Ni
nanoparticles\cite{Herrero}. We show an example of a
trigonal-zigzag-nanodisk array sharing one zigzag edge as in Fig.8(a). We
show the corresponding band structure in Fig.8(b). It is intriguing that
there are $N$-fold degenerate perfect flat bands in the nanodisk with size $%
N $. This fact is confirmed by Leib's theorem. Each nanodisk has spin $N/2$
and makes ferromagnetic coupling between two nanodisks. In the same way we
can make two-dimensional nanodisk arrays. It is to be emphasized that they
show ferromagnetism, and not quasiferromagnetism, though they are made of
nonmagnetic materials. The perfect flat band will be robust even when
electron interactions are introduced.

\begin{figure}[t]
\centerline{\includegraphics[width=0.48\textwidth]{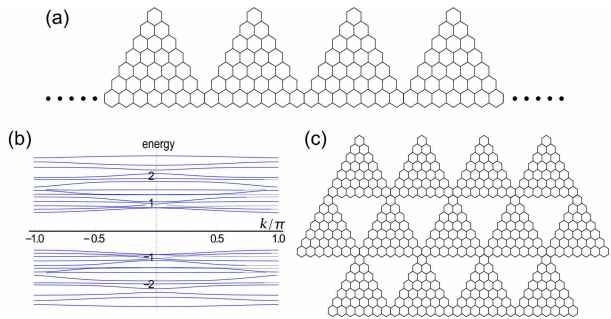}} \label{FigFChain}
\caption{
(a) Illustration of
one-dimensional nanodisk arrays. (b) Band structure of one-dimensional
nanodisk arrays.(c) Illustration of two-dimensional nanodisk arrays.}
\end{figure}

\textbf{Nanomechanical switch:} We construct a nanomechanical switch
contacting two graphene trigonal corners [see Fig.9(a)]. We assume the angle
between two corners is $\theta $. This angle is tuned by an external
mechanical force. The carbon-skeleton structure is made of $\sigma $-bonds,
and is very rigid except for this rotational degree of freedom.

\begin{figure}[t]
\centerline{\includegraphics[width=0.39\textwidth]{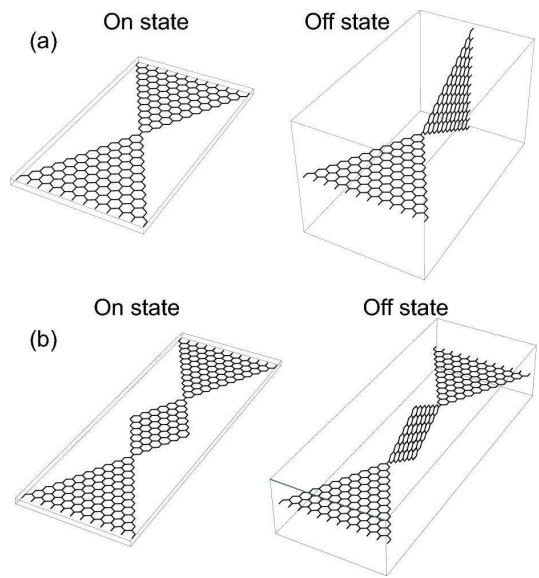}} \label%
{FigCornerStream}
\caption{
(a) Illustration of a
nanomechanical junction made of two triangular graphene corners. There are
on and off states, switched by a rotation between two cornes. (b)
Illustration of a nanomechanical rotator made of two corners and a central
rhombus rotating rather freely. }
\end{figure}

The conductance is determined by the overlap integral of $\pi $-electrons
between two corners, which is given by%
\begin{equation}
\left\vert \left\langle p_{z}\cos \theta +p_{y}\sin \theta \right.
\left\vert p_{z}\right\rangle \right\vert =\cos ^{2}\theta .
\end{equation}%
When the two planes are parallel ($\theta =0$), the overlap takes the
maximum value and $\pi $-electrons can go through the contact. This is the
on state. When the two planes are orthogonal ($\theta =\pi /2$), the overlap
takes the minimum value and $\pi $-electrons can not go through the contact.
This is the off state. The angle is changed nanomechanically. The system
acts as a nanomechanical switch. It could detect the angle very sensitively
and be useful for detect nanomechanical oscillations.

By connecting two nanomechanical junctions, we can construct a
nanomechanical rotator [Fig.9(b)], where two corners are suspended
mechanically while the central rhombus rotates rather freely. This structure
may be useful to detect molecular dynamics. When molecules contact the
rotator, they are detected by rotating it and changing the resistance
between the two corners.

\textbf{Peierls instability:} We connect two triangular corners with a
zigzag nanoribbon [see Fig.10(a)]. This structure has already been
manufactured experimentally\cite{Herrero}. Polyacetylene has the Peierls
instability: Carbons with conjugate bonds are spontaneusly deformed into
alternating single and double bonds. We expect the Peierls instability to
occur also in graphene nanoribbons [Fig.10(b)], because grapahene nanoribbon
is a natural extension of polyacetylene\cite{EzawaRibbon}. (We denote the
width of a nanoribbon by $W$ with $W=0$ for polyacetylene and $W=1$ for
polyacene.) However, as we now show, the Peierls instability will not occur
in nanoribbons with the width $W>2$. On the contrary, it is possible to
induce the Peierls transition manually by streaching a nanoribbon. Making an
advantage of this property, we propose a nanomechanical switch sensor.  

\begin{figure}[t]
\centerline{\includegraphics[width=0.44\textwidth]{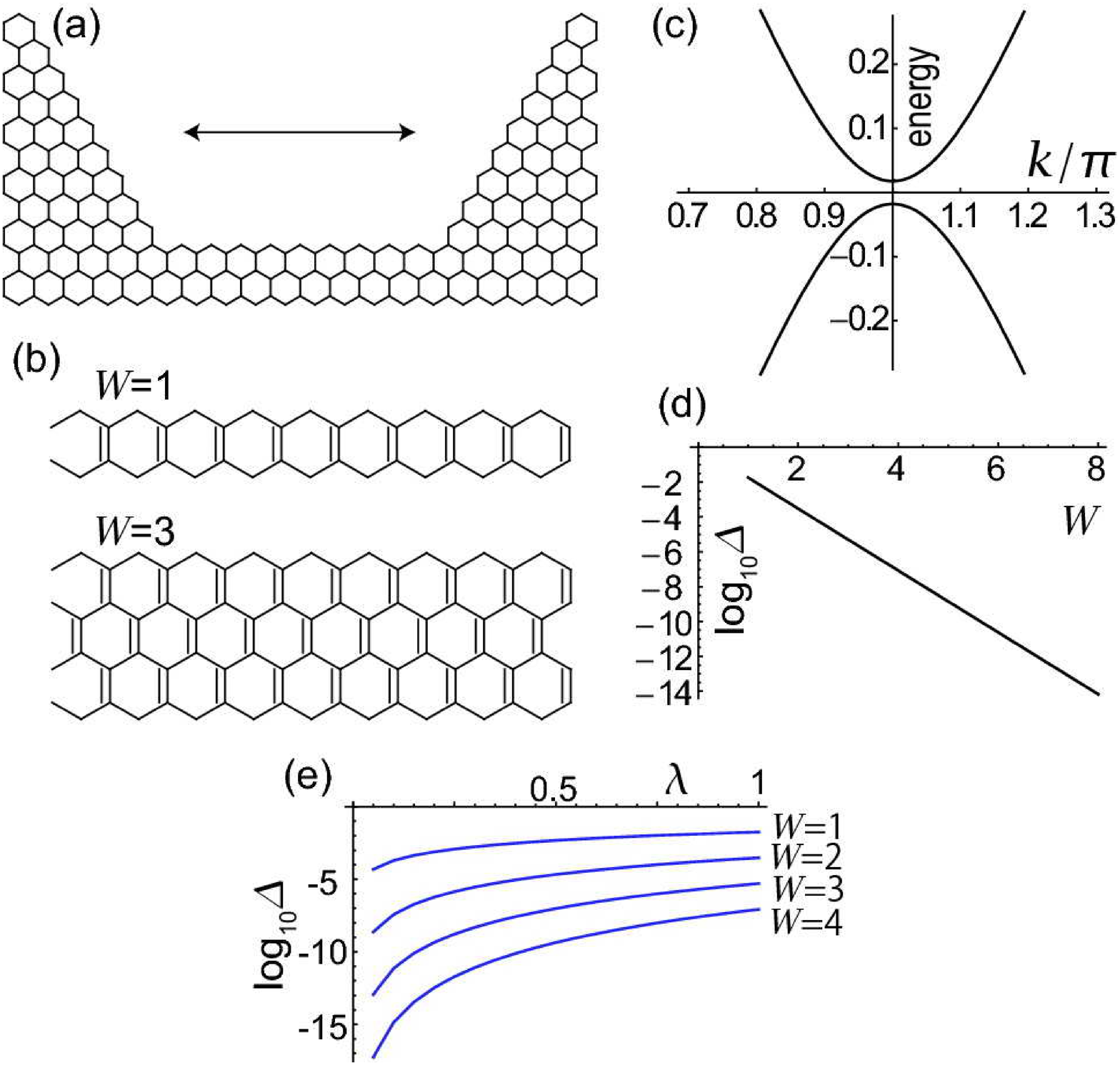}} \label%
{FigPeierls}
\caption{
(a) Illustration of a configuration where a
single ziazag nanoribbon is connected by two triangular graphene corners.
(b) Illustration of the Peierls instability of graphene nanoribbons with $%
W=1 $ and $W=3$, where $W$ represents the width. Single and double bonds
alternatingly appears. (c) The band structure around $k=\pi $ of a polyacene
with the Peierls instability. (d) The logarithm plot of the band gap $\Delta 
$, $\log _{10}\Delta /t$, of nanoribbons with various $W$. (e) The logarithm
plot of the band gap $\Delta $, $\log _{10}\Delta /t$, of nanoribbons as a
function of the applied force $\delta $ for various $W$.}
\end{figure}

The model Hamiltonian is the Su-Schrieffer-Heeger-like model\cite{SSH},%
\begin{equation}
H_{0}=\sum_{\left\langle i,j\right\rangle }t_{ij}c_{i,\sigma }^{\dagger
}c_{j,\sigma }.
\end{equation}%
In this model, when the Peierls instability occurs, the transfer integral
takes two values corresponding to the single and double bonds, and otherwise
it takes one value corresponding to the conjugate bond. We are able to
calculate the gap analytically in the case of polyacene ($W=1$), where the
band structure around $k=\pi $ is given by%
\begin{equation}
\varepsilon \left( k_{x}\right) =\pm \frac{1}{2}\left( -t_{2}+\sqrt{%
4t_{1}^{2}+5t_{2}^{2}+8t_{1}t_{2}\cos k_{x}}\right) .
\end{equation}%
The band gap is determined by substituting $k_{x}=\pi $ as%
\begin{equation}
\Delta =2\varepsilon \left( \pi \right) =\left( -t_{2}+\sqrt{%
4t_{1}^{2}+5t_{2}^{2}-8t_{1}t_{2}}\right) \sim 4\delta t_{1}\delta t_{2}.
\end{equation}%
We show $\varepsilon \left( k_{x}\right) $ in Fig.10(c) by taking the values 
$t_{1}=t-\delta t_{1}=0.94t$ for single bonds and $t_{2}=t+\delta t_{2}=1.08t
$ for double bonds. We have carried out a numerical estimation of the band
sturacture for $W\geq 2$. A tiny gap $\Delta $ opens at $k=\pi $. The
logarithm plot of the band gap $\Delta $ as a function of the width $W$ is
shown in Fig.10(d). The gap decreases exponentially as a function of the
width, $\propto 10^{-1.7W}$. The gap of the case $W=1$ (polyacene) is 49meV,
and that of the case $W=2$ is 0.8meV. 

We next compare the energy gain from this gap with the energy cost from the
elastic energy of a lattice deformation. The ground-state energy difference
between distorted and undistorted structures is very tiny in polyacene\cite%
{Longet}. On the other hand, the elastic energy cost is proportional to the
width $W$, and hence the elastic energy cost becomes larger than the gap
energy gain for wider nanoribbons. The Peierls instability will not occur
spontaneously in nanoribbons with $W>2$.

We have also estimated how the band gap $\Delta $ depends on the external
force. The transfer integrals change proportionally to the external force.
For simplicity we have set $t_{1}=\left( 1-0.06\lambda \right) t$ and $%
t_{2}=\left( 1+0.08\lambda \right) t$. It takes about $10$GPa for
deformation $0.01t$ for graphene, while it takes about $1$GPa for
deformation $0.01t$ for narrow graphene nanoribbons\cite{Zhao}. We show the
logarithm plot of the $\lambda $ dependence of the band gaps in Fig. 10(e).

We propose an application of the above system. By streching a nanoribbon
along the ribbon direction, horizontal bonds are stretched and vertical
bonds are shrinked. The resultant structure is resemble to the deformed
structure induced by the Peierls transition. We may call it a strain induced
Peierls transition. When we stretch this structure, the band gap opens and
the conductance at the zero energy becomes zero by the strain induced
Peierls-transition. On the other hand, without the external mechanical
force, nanoribbons with $W>2$ is gapless and the system is conductive. We
can switch the conductance from on to off by stretching the system. The
system acts as a nanomechanical switch sensor detecting nanoscale
displacement.

\section{Conclusions}

A nanodisk can be used as a spin filter just as in a metal-ferromagnet-metal
junction. A novel feature is that the direction of the spin can be
controlled by external field or spin current. We have newly proposed
nanodisk arrays and nanomechanical switch. These nanodisk-nanoribbon complex
structure will open a new field of nanoelectronics, spintronics and
nanoelectromechanics purely based on graphene.

\bigskip \noindent\textbf{Acknowledgements} \bigskip

I am very grateful to Y. Takada, H. Tsunetsugu, B.K. Nikolic and N. Nagaosa
for fruiteful discussions. This work was supported in part by Grants-in-Aid
for Scientific Research from the Ministry of Education, Science, Sports and
Culture No. 22740196 and 21244053.

\end{document}